\documentclass[superscriptaddress,preprint,amsmath,amssymb]{revtex4}
\usepackage{graphicx}  
\usepackage{latexsym}
\usepackage{dcolumn}   
\usepackage{bm}        
\usepackage{amssymb}   
\usepackage{textcomp}
\usepackage{amsmath}
\usepackage[dvips]{color}
\usepackage{soul}
\usepackage{float}

\begin{document}

\title{Ionic liquids make DNA rigid}

\author{Ashok Garai}
\email[Corresponding author: ]{ashok.garai@gmail.com}
\address{Centre for Condensed Matter Theory, Department of Physics, Indian Institute of Science, Bangalore 560012, India}
\affiliation{Department of Physics, The LNM Institute of Information Technology, Jamdoli, Jaipur 302031, India},
\author{Debostuti Ghoshdastidar}
\address{Bhupat and Jyoti Mehta School of Biosciences, Department of Biotechnology, Indian Institute of Technology Madras, Chennai 600036, India.}
\author{Sanjib Senapati}
\affiliation{Bhupat and Jyoti Mehta School of Biosciences, Department of Biotechnology, Indian Institute of Technology Madras, Chennai 600036, India.}
\author{Prabal K. Maiti} 
\affiliation{Centre for Condensed Matter Theory, Department of Physics, Indian Institute of Science, Bangalore 560012, India}

\date{\today}

\begin{abstract}
Persistence length of dsDNA is known to decrease with increase in ionic concentration of the solution. In contrast to this, here we show that persistence length of dsDNA increases dramatically  as a function of ionic liquid (IL) concentration. Using all atomic explicit solvent molecular dynamics simulations and theoretical models we present, for the first time, a systematic study to determine the mechanical properties of dsDNA in various hydrated ionic liquids at different concentrations. We find that dsDNA in $50$wt\% ILs have lower persistence length and stretch modulus in comparison to $80$wt\% ILs. We further observe that both persistence length and stretch modulus of dsDNA increase as we increase the ILs concentration. Present trend of stretch modulus and persistence length of dsDNA with ILs concentration supports the predictions of the macroscopic elastic theory, in contrast to the behavior exhibited by dsDNA in monovalent salt. Our study further suggests the preferable ILs that can be used for maintaining DNA stability during long-term storage.
\end{abstract}

\pacs{}
\keywords{DNA, Ionic liquids, Persistence length}
\maketitle

\section{Introduction}
In a eukaryotic cell DNA is strongly bent on histone octamer proteins and forms a beads on string structure \cite{alberts08}. In this way it is tightly packed inside the nucleus. All active biological processes like transcription, replication, recombination, and DNA packaging occur at short length scale of DNA ($<50$ nm). Thus study of various mechanical properties at this length scale is very important. Technological advancement now makes it possible to measure various mechanical properties of such short DNAs \cite{fenn08, yuan08, bomble08, noy12, maiti13, mazur14, maaloum14, wu15, garai15}. Recent experiments \cite{fenn08, yuan08, mazur14, maaloum14, cloutier04, wiggins06, vafabakhsh12} and theoretical as well as molecular dynamics (MD) studies \cite{bomble08, noy12, maiti13, wu15, garai15} suggest that short DNAs are more flexible than kilo base pairs (kbps) of DNA \cite{baumann97}. These studies indicate that long DNAs (kbps) can be well described by Worm Like Chain (WLC) model \cite{marko95, kumar10} with a persistence length of $~ 50$ nm. The WLC model assumes that long DNA can be treated as a semi flexible polymer having a quadratic form for bending deformation energies. Recently measured mechanical properties of short- and intermediate-length DNAs strongly challenge the use of WLC model \cite{fenn08, yuan08, wiggins06, vafabakhsh12}.  In an earlier study, we \cite{garai15} reported that persistence length and stretch modulus of short DNAs depend on DNA lengths as well as on monovalent salt concentrations. Persistence length decreases with increase in salt concentrations whereas stretch modulus increases with increase of salt concentrations.

In recent years, DNA has been used as a biomaterial with applications in varied fields ranging from therapeutics \cite{patil05, liu03} to molecular computing \cite{goldman13, normile02}. The interest of DNA in nanotechnology has led to employing DNA for design of nanoscale objects of varying topologies \cite{maiti16, maiti17, maiti15, maiti06, maiti04, liu04, liu09, yamada05}. However, these applications are often hindered by the structural and chemical instability of DNA. The structure of DNA depends on various conditions like temperature, pH, salt strengths etc. While aqueous solution was considered as a reasonably good solvent in this regard, DNA was found to be chemically unstable in it during long-term storage \cite{cheng92, lindahl72}. Moreover, use of aqueous solutions in small-volume technology is difficult since water vaporizes immediately under open-air conditions or at high temperatures \cite{sugimoto14}. Use of non-aqueous and mixed solvents is also limited as DNA loses its structure in these media \cite{bonner00, hammouda06}. Thus finding an appropriate solvent where DNA remains soluble without losing its structure has remained challenging. Recently, a new class of organic solvents that are composed of ions, which are liquid at room temperature, called ionic liquids (ILs), have exhibited tremendous potential as a medium for bimolecular dissolution and stability \cite{sugimoto14, macfalrane10, senapati12,liu12}. For details regarding the chemical and molecular structure of ILs, the reader is referred elsewhere \cite{hayes15}. In a pioneering study, MacFarlane and coworkers \cite{macfalrane10} have shown that DNA retains its structure in various ILs during long-term storage at room temperature. Using molecular dynamics simulations, Senapati et al. \cite{senapati12} elucidated the atomistic basis for such exceptional stability exhibited by DNA in ILs. Their results suggest that ILs disrupt the water cage around the duplex, thereby preventing hydrolytic damage during long-term storage of DNA. More interestingly, ILs were found to interact with the minor and major grooves and the phosphate backbone of the DNA, conferring additional structural stability on the biomolecule.

Considering the potential use of ILs as the next-generation solvent medium for DNA, and their unique ability to confer exceptional stability on DNA structure, it is natural to ask what is the microscopic origin of this enhanced stability? What prevents the hydrolytic damage of DNA in IL solution. We hypothesize that enhanced rigidity of DNA in IL makes  its interaction and/or binding to different metabolites or DNase unfavorable and hence prevents degradation. To test this hypothesis, in this letter we have investigated various mechanical properties, like persistence length and stretch modulus, of dsDNA in presence of a range of hydrated ILs at different concentrations. The use of hydrated ILs, as opposed to neat ILs, is prompted by the earlier report of Senapati {\it et al.} \cite{senapati12} where a solute amount of water was always found to be retained in the hydration shell of DNA. We observe that DNA behaves like a rigid elastic rod in hydrated IL, contrary to its nature in monovalent salt. Our study further indicates that mechanical properties of DNA in different ILs could be the determinant for choosing the appropriate IL for long-term DNA storage. 

\begin{figure}[tb]
\centering
\includegraphics[width=0.69\columnwidth]{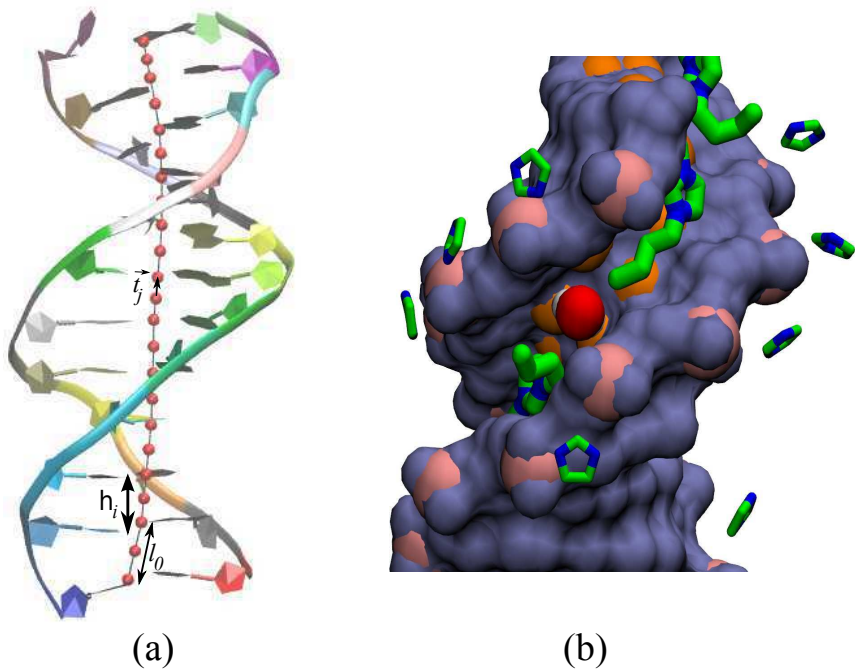}
\caption{(color online). (a) DDd with its central axis traced using Curves+ \cite{lavery09}. Successive heights between base pairs are $h_i$ as shown in figure and $\vec{t}_j$ is the local tangent vector at $j$. (b) Snapshot of DNA solvated in $80$wt\% {\it [bmim][lactate]} solution. The imidazolium cation head group interacts with the phosphate backbone (pink) of the DNA. The cation also penetrates deep into the minor groove where its polar head group and hydrophobic tail establish electrostatic and van der Waals interactions, respectively, with the base atoms (orange). While IL penetration severely dehydrates the DNA solvation shell, solute amount of water (red) necessary for DNA structural integrity is retained in the hydration spine of the minor groove. The parts of the [bmim] cation specifically interacting with different regions of the DNA have been shown - imidazolium headgroups interact with the backbone while both the imidazolium head group and the alkyl tails penetrate into the grooves.}
\label{fig1}
\end{figure}

\section{Methods}
\subsection{Simulation Details}
We have performed a series of all atom MD simulations of Dickerson-Drew dodecamer (DDd, $5^{'}-d-(CGCGAATTCGCG)_2-3^{'})$  DNA \cite{senapati12,perez07} in binary mixtures of the widely studied imidazolium and cholinium classes of ILs with water. Chemical structures of IL cations and anions used in the present study are shown in Fig. S1. ILs belonging to the imidazolium class include $1-$ethyl$-3-$methyl imidazolium acetate {\it ([emim][acetate])}, $1-$butyl$-3-$methyl imidazolium acetate {\it ([bmim][acetate])}, $1-$butyl$-3-$methyl imidazolium lactate {\it ([bmim][lactate])} and $1-$butyl$-3-$methyl imidazolium nitrate {\it ([bmim][nitrate])}, while {\it [choline][acetate], [choline][lactate]} and {\it [choline][nitrate]} ILs were selected from the cholinium class. Additionally, a control simulation of DNA in water was also carried out. The water:IL ratios in the simulations were chosen based on the concentrations of ILs in the experiments of MacFarlane and co-workers \cite{macfalrane10}. DNA dodecamer was placed at the center of a cubic box of length $6.5$ nm with periodic boundary conditions in all directions. $22$ Na$^{+}$ ions and $8000$ water molecules were added to the system. The box size was chosen such that any DNA atom is at least $1.5$ nm away from the box surface, preventing unwanted interactions with its image in translated unit cells. For the IL systems, the equilibrated dodecamer with four layers of surrounding water molecules from aqueous system was placed into a cubic box of length $6.5$ nm. This box was then filled with requisite number of IL pairs and water to reach the desired IL concentrations. The systems were kept neutral by adding $22$ Na$^{+}$ ions. Further details of the simulated systems are presented in Table {\it SI} in the supplementary material.

Amber parm99 force field \cite{kollman99} with parmbsc0 corrections \cite{orozco07} and TIP3P water model \cite{klein83} were adopted to represent the interaction potentials of DNA and water, respectively. The interaction potentials, used for imidazolium- and choline-based ILs were adopted from literature, which were developed within the spirit of OPLS-AA/AMBER framework \cite{lopes06, somisetti09}. Simulations were carried out at $298$ K and $1$ atm pressure. Simulations were performed in NPT ensemble using the Berendsen thermostat and barostat. Both the thermostat and barostat relaxation times were set to $0.5$ ps. The calculation of long-range Coulomb interaction was performed employing the full Ewald summation technique. $15$ \AA{} cut-off was used to compute real space part of the Ewald sum and Lennard-Jones interactions. SHAKE was used to constrain bond lengths involving hydrogens atoms. A set of minimization and equilibration runs of the starting structures was performed to remove the initial bad contacts. Simulations were performed for each system until the potential energy and system volume ceased to show any systemic drift. Subsequently, a production phase of $100$ ns was performed using SANDER module of Amber12 \cite{cheatham13}.

\subsection{Analytical tools}
\subsubsection{Calculation of the persistence length $(l_p)$}
DNA in its discrete description can be described with equal segments of $l_0$. Orientational correlation between two tangent vectors $\vec{t}_j$ and $\vec{t}_{j+n}$ (see Fig.~\ref{fig1}(a)) separated by a distance $d$ along the DNA axis represents the persistence length $(l_p)$ of the DNA. Thus persistence length is a characteristic length over which the tangent-tangent correlations die off. In other words, it also measures the local bending between the two tangents with respect to the local axis $\hat{z} = \vec{t}_j \times \vec{t}_{j+n}$ with amplitude $cos\theta_j = \vec{t}_j \cdot \vec{t}_{j+n}$, where $\theta_j$ is the bending angle at $j$. In Kratky-Porod model \cite{kratky49} it assumes that DNA resists to bending deformations, which is captured through an elastic bending energy defined as 
\begin{equation}
E_{K-P} = \frac{l_p}{l_0} \sum^{N-1}_{j=1}(1-cos\theta_j)
\label{eq-ekp}
\end{equation}

where $l_p = \beta \kappa$, $\beta= 1/k_BT$, $k_B$ is the Boltzmann constant, $T$ is the temperature and $\kappa$ is the bending modulus. Small fluctuations in $\theta$ can lead to a bending probability distribution and that can be approximated to Gaussian distribution as follows.
\begin{equation}
P(\theta) = \sqrt{\frac{\beta \kappa}{2\pi L_0}} e^{-\frac{\beta \kappa}{2L_0}\theta^2}
\label{eq-pthe}
\end{equation}
where $L_0 = \sum_j h_j$ is the contour length of the DNA (see Fig. 1). For small bending angles Eq.~(\ref{eq-pthe}) can further be simplified and is given by
 \begin{equation}
lnP(\theta) = -\frac{l_p}{L_0}(1-cos\theta) + 0.5ln(\frac{\beta \kappa}{2\pi L_0})
\label{eq-pt}
\end{equation}
A plot between $lnP(\theta)$ vs. $(1-cos\theta)$ would give rise to a straight line with a descending slope. Thus fitting simulation/experimental data with Eq.~(\ref{eq-pt}) and from the slope one can estimate the persistence length of the DNA. The method followed here is similar in spirit to the procedure reported in different earlier publications. \cite{maiti16, wu15, garai15, wiggins06, maiti13, zoli18, zoli16, zoli2016, mazur07, mazur2014, marko15}.

\subsubsection{Calculation of the stretch modulus $(\gamma)$}
A DNA can be modeled as an elastic rod. In the elastic rod model \cite{bustamante94, baumann97}, the deformation in DNA is restricted to the linear elastic regime. Assume that the DNA has a time averaged contour length $L_0$ whereas the instantaneous length of the DNA is $L$. In presence of small perturbation along its length produces a restoring force $F$. This model assumes that this $F$ increases linearly with the instantaneous fluctuations $(L-L_0)/L_0$. Thus $F = - \gamma(L-L_0)/L_0$ , where $\gamma$ is the stretch modulus of the ds-DNA. The associated free energy change becomes $E(L) = \gamma(L-L_0)^2/2L_0$. Substituting the expression of $E(L)$ into Boltzmann's formula we obtain the following expression for the probability of ds-DNA with contour length $L$	

\begin{equation}
P(L) = \sqrt{\frac{\beta \gamma}{2\pi L_0}} e^{- \frac{\beta \gamma}{2L_0}(L-L_0)^2}
\label{eq-pl}
\end{equation}

Eq.(\ref{eq-pl}) looks like Gaussian function with a mean contour length $L_0$, and variance $L_0/\beta \gamma$. Fitting experimental/simulation data with Eq.(\ref{eq-pl}) one can estimate the stretch modulus.

\section{Results and Discussion}
DDd was simulated in water and several IL/water binary mixtures as listed in Table {\it SI}. For all solutions, two different IL:water compositions were selected, $80:20$ and $50:50$, to study the effect of IL concentration on the mechanical properties of DNA. The structural stability of DNA in IL/water binary mixtures was probed by analyzing its root mean square deviation (RMSD) from the native crystal structure throughout the simulation trajectory (see figures S2 and S3 in the supplementary material). Time evolution of RMSD showed that irrespective of the IL type or concentration, the B-form of DNA was stable, indicated by a structural deviation of $< 2.5$ {\AA} from the canonical B-DNA crystal structure during the production run of $100$ ns. Notably, in this and earlier studies we have observed that the dynamics of DNA is considerably dampened in presence of ILs, due to the high viscosity of the IL/water medium. However, computation of DNA helical parameters in all IL/water mixtures has shown that DNA samples almost equivalent conformational space in IL/water mixtures as in aqueous solutions. For example, Figure S4 in the supplementary material presents the rise and twist values sampled by the DDd in presence and absence of [choline][acetate] IL. Evidently, changes in these helical parameters observed during the simulation period are comparable in both solvent media, indicating that albeit slowly DNA does sample the B-DNA conformational space in the binary mixtures as it does in water. Many DNA conformational dynamics such as bending, twisting, backbone dynamics and sugar puckering happens in a timescale of $100$ ns or faster \cite{galindo14}. DNA dynamics associated with the base pair opening happens in the timescale of millisecond. So our ns long DNA dynamics will probe majority of the DNA conformational change in a satisfactory manner except dynamics associated with base pair opening. Thus DNA persistence length calculated using bending angle distribution from the $100-200$ ns long dynamics should be realistic and statistically significant.

\begin{figure}[tb]
\centering
\includegraphics[width=0.98\columnwidth]{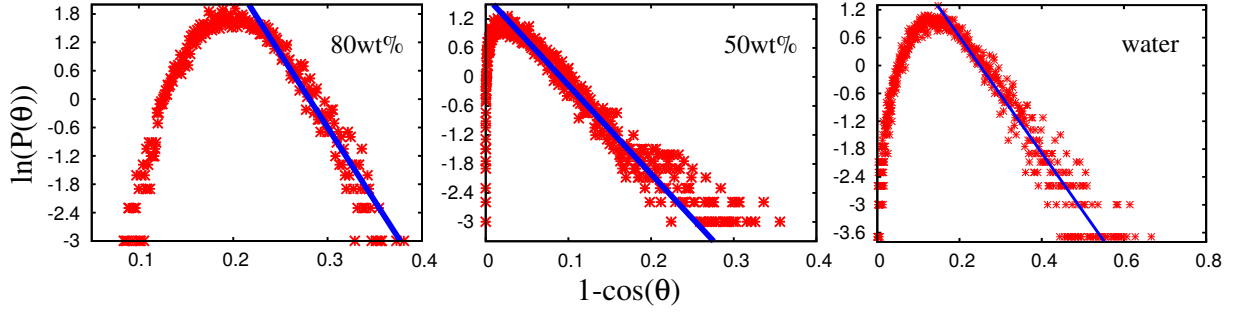}
\caption{(color online). Probability distributions of bend angles for DDd at various {\it [choline][acetate]} concentrations. Discrete data points (red) are obtained from our MD simulations. Persistence lengths at respective cases are calculated by fitting the slope with Eq.~(\ref{eq-pt}) and are shown by straight line (blue).}
\label{fig2}
\end{figure}

To determine the persistence length ($l_p$) of DNA in presence of ILs, the bending angle distributions of DDd solvated in various IL/water binary mixtures at different concentrations were deduced. Fig.~\ref{fig2} illustrates log of bending angle distribution as a function of $(1-cos\theta)$ for {\it [choline][acetate]} at different IL concentrations - $80$wt\%, $50$wt\% and $0$wt\% . Discrete data points (in red) were obtained from our computer simulations.  The $l_p$ of DNA in the respective systems was calculated by fitting the data with Eq. (\ref{eq-pt}), as represented by solid lines (in blue) in the figure. The $l_p$ of DDd in presence of other ILs at different concentrations are listed in Table \ref{Table1}. These values are further plotted in Fig. S5 of the supplementary material. As evident from the figures, the $l_p$ of DDd is strongly modulated by IL concentration as well as IL class. Interestingly, for majority of ILs the $l_p$ of DNA increases with increasing IL concentration. Such increase in $l_p$ with increasing ionic strength is contrary to the behavior exhibited by DNA in inorganic salt solutions, where the $l_p$ of DNA decreases with increasing salt concentration. It is well known that inorganic cations neutralize the negatively charged phosphate backbone of DNA thereby increasing the bendability of the duplex. As a result, high monovalent salt concentration leads to soft DNA with lower $l_p$. Conversely, IL cations not only bind to the phosphate backbone but also establish strong electrostatic and hydrophobic interactions with the DNA grooves. As shown in Fig.~\ref{fig1}(b) the imidazolium cation penetrates deep into the minor groove of the DNA duplex where it makes electrostatic and hydrophobic interactions with the base and sugar moieties  \cite{senapati12}. Such strong IL-DNA interactions confers rigidity on the DNA structure causing the observed increase in $l_p$. Notably, while this increase is marked in $80$wt\% hydrated IL solution, the $l_p$ of DNA in $50$wt\% solution shows only slight increase from that in water. In the $50$wt\% IL solution, the DNA groove contains significant water with respect to IL. An earlier study has shown that in presence of larger amounts of water, IL/water binary mixtures exhibit more water-like than IL-like properties \cite{senapati17}. As a result, the DNA $l_p$ at this concentration is found to be similar to that in water medium. On the other hand, at $80$wt\%, DNA grooves are populated primarily by IL cations and the difference in $l_p$ with respect to neat water system becomes evident. Another interesting fact is the higher $l_p$ of DNA in {\it [bmim]}-based ILs compared to {\it [choline]} or {\it [emim]} systems at $80$wt\% (Table \ref{Table1}). While all three IL cations bind with the DNA minor groove (Fig.~\ref{fig1}(b)), they exhibit different modes of interaction with the groove atoms. As shown in the Fig. \ref{fig1}(b), the planar imidazolium head group and long hydrophobic tail of the {\it [bmim]} cation enable easy accessibility into the depths of the minor groove where the cation establishes electrostatic and van der Waals interactions with the base atoms. Conversely, the bulky NMe$_3$ headgroup of {\it [choline]} doesn't allow it to penetrate deep into the groove, resulting in comparatively weaker binding. The shorter ethyl chain of {\it [emim]} cation also weakens its interaction with the DNA groove. Thus, stronger the interactions of DNA with IL, more rigid the DNA duplex becomes. Notably, unlike other ILs, the $l_p$ of DNA solvated in {\it [bmim][nitrate]}/water mixtures exhibited values close to the control water system irrespective of the IL concentration. In agreement with an earlier report, among all ILs {\it [bmim][nitrate]} was found to strip least amount of water from the DNA solvation shell \cite{senapati12}. This is plausibly because {\it [bmim]} remains strongly paired with the {\it [nitrate]} anion, preventing the cation from penetrating into the narrow minor groove of DNA. The significant amount of water retained in the solvation shell of DNA dissolved in the {\it [bmim][nitrate]} system results in the low $l_p$ of the duplex in this system. Therefore, it can be clearly surmised from the above findings that the level of hydration of DNA is inversely related to its persistence length in IL/water binary mixtures.   

\begin{table*}
\caption{Mechanical properties of DDd in various ILs at different concentrations.}
\centering
\resizebox{\columnwidth}{!}{%
\begin{ruledtabular}
\begin{tabular*}{\textwidth}{@{\extracolsep{\fill}}lllll}
 DDd in & Persistence length  &Persistence length  & Stretch modulus  & Stretch modulus  \\
  & (nm) at $80$wt\% & (nm) at $50$wt\% & (pN) at $80$wt\% & (pN) at $50$wt\% \\
 \hline
 $[bmim][acetate]$ & $248.74 \pm 6.8$ & $58.34 \pm 1.53$ & $3926.24 \pm 124.02$  & $981.87 \pm 86.93$ \\
 $[bmim][lactate]$ & $131.42 \pm 4.54$ & $61.77 \pm 3.11$ & $2980.85 \pm 67.84$ & $3653.42 \pm 61.57$ \\
 $[bmim][nitrate]$ & $49.67 \pm 1.79$ & $35.10 \pm 1.38$ & $2771.38 \pm 52.79$ & $1085.98 \pm 25.84$ \\
 $[choline][acetate]$ & $126.35 \pm 4.16$ & $74.03 \pm 0.99$ & $8402.44 \pm 114.28$ & $1640.32 \pm 36.59$ \\
 $[choline][lactate]$ & $113.49 \pm 3.73$ & $73.38 \pm 1.38$ & $6324.84 \pm 86.57$ & $2579.09 \pm 27.42$ \\
 $[choline][nitrate]$ & $105.21 \pm 1.90$ & $55.83 \pm 0.80$ & $4457.95 \pm 61.45$ & $3127.66 \pm 93.42$ \\
 $[emim][acetate]$ &  $99.91 \pm 2.33$ & $49.18 \pm 1.26$ & $5951.72 \pm 175.26$ & $1999.64 \pm 50.98$ \\
  water (no IL) & \qquad $49.89 \pm 0.77$ &  & \qquad $1096 \pm 32$  &  \\
   \end{tabular*}
  \end{ruledtabular}
  }
\label{Table1}
\end{table*}

\begin{figure}[tb]
\centering
\includegraphics[width=0.65\columnwidth]{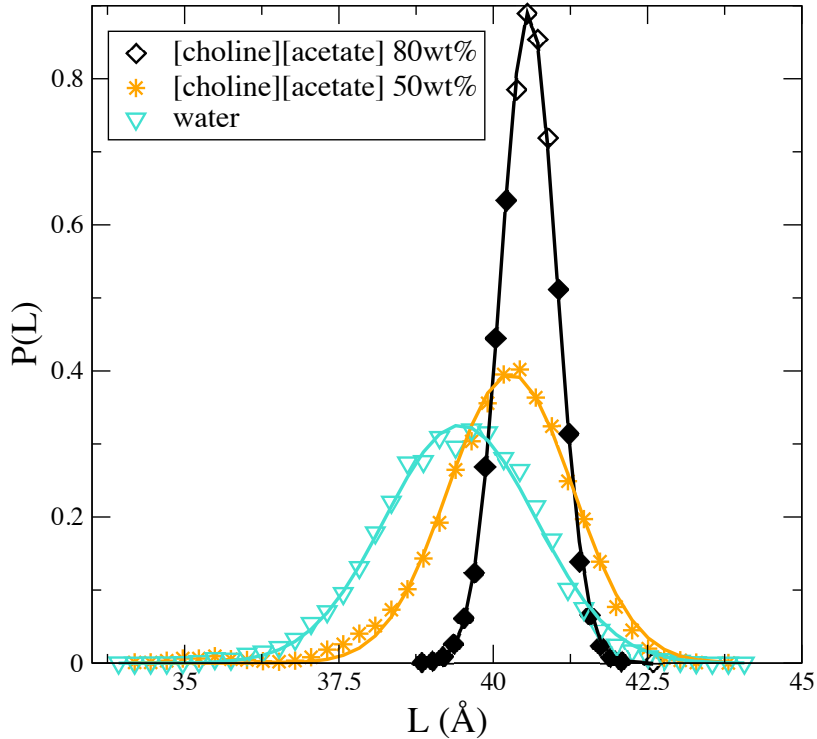}
\caption{(color online). Contour length distribution for DDd at different ILs concentration and at no IL (water) solution. Discrete data points are obtained from our MD simulations and the respective solid lines are obtained by fitting the data points with Eq.~(\ref{eq-pl}).} 
\label{fig3}
\end{figure}

Using our MD simulation data, we further calculated the contour length distribution $(P(L))$ of DDd at various IL solutions. For all the considered ILs we observed (not shown) that the average contour length of the DDd in $50$wt\% is comparable in absence and presence of IL. This is in consensus with the groove-binding mode of IL with DNA, which, as opposed to intercalation, does not alter the total length of the DNA. Contour length distributions of DDd in {\it [choline][acetate]} at different concentrations and in water are illustrated in Fig. \ref{fig3}. Discrete data points were obtained from simulations and the solid lines were obtained by fitting these data points with Eq.~(\ref{eq-pl}). Interestingly, the contour length distribution of DDd in $80$wt\% IL solution is narrower than the other two cases. This indicates that with increasing IL concentration, DNA explores lesser number of structural variants due to the increased rigidity of the IL-bound DNA structure, in agreement with our previous observation. Stretch modulus values of DDd in the respective IL environments were calculated by fitting the discrete data with Eq. (\ref{eq-pl}), as represented by solid lines in the Fig. \ref{fig3}. Hence, we compared the $\gamma$ values of DDd in various ILs at different concentrations and listed them in Table \ref{Table1}. The respective $\gamma$ values are also plotted in Fig. S6 in the supplementary material. We observe that with addition of IL, the stretch modulus of DDd increases with increasing IL concentration. A larger stretch modulus is indicative of a stiffer DNA in presence of ILs. We observe that stretch modulus and persistence length of DDd in different ILs increase with the increase of the respective IL concentrations, which is unlike to the situation when DNA in monovalent salt\cite{baumann97, garai15}. Present trend supports macroscopic elastic theory \cite{baumann97, garai15}, which predicts these two quantities to be directly proportional to each other. Yet another interesting fact evident from Table \ref{Table1} is the several-fold higher stretch modulus of DDd in presence of {\it [choline]}-based ILs compared to that in {\it [bmim]}-based ILs in $80$wt\% binary mixtures. Recall that this trend is opposite to the higher $l_p$ of DNA observed in {\it [bmim]}-containing binary mixtures. Presence of all cations are known to increase the stretch modulus of DNA by neutralizing the DNA backbone and lowering intra-molecular electrosatic repulsion. However, unlike {\it [bmim], [choline]} ILs interact with greater affinity with the DNA backbone owing to the localized positive charge on the NMe$_3$ headgroup of the choline cation. As a result, it requires larger force to stretch [choline]-bound DNA, leading to higher values of stretch moduli in these systems. Moreover, as observed by us and others, in presence of {\it [choline]}-ILs, higher amount of water is retained in the DNA solvation shell compared to {\it [bmim]}-ILs. The -OH group of the choline cation is known to interact with water and form a complex cation-water hydrogen bonded network \cite{konstantin99}. Such interactions in the solvation shell of DNA dissolved in {\it [choline]}-ILs plausibly makes the hydration layer quite rigid, thereby making it difficult to stretch the DNA. Notably, a similar observation was made in an earlier study where the presence of an ordered hydration network was found to increase the elastic stretch modulus of a self-assembled DNA monolayer \cite{carmen17}. It is worth mentioning here that the knowledge of DNA flexibility or deformability at the base step level also can give important insight into their biological action. Earlier Lankas et. al. \cite{lankavs03} have investigated the elastic constants for each base pair step in aqueous environment. Following their analysis we also investigate the local stiffness parameters of the DNA in two different concentrations of [choline][acetate] (not shown). We find that the stiffness parameters at the base pair level in the above mentioned IL vary based on the base sequences and they get modified with the IL-phosphate interactions. These observations further suggest that DDd becomes more rigid with the increase of IL concentration. A detailed analysis of the local stiffness parameters of the DNA in different IL environments is currently under our investigation.

\section{Conclusions}
Presence of hydrated ionic liquid stabilizes the B-conformation of the DDd \cite{macfalrane10, senapati12}. The atomistic basis for long-term stability of DDd in hydrated liquids is also explored by studying various mechanical properties of DNA via MD simulations along with various theoretical models. The results suggest that mechanical properties of DDd depend not only on IL solution concentration but also on different types of ILs. Persistence length of DDd in an $80$wt\% IL solution is higher than that in a $50$wt\% IL solution. High density of IL cations in the DNA solvation shell reduces the inter-strand phosphate repulsion but increases the electrostatic repulsion in the intra-strand. Further binding of cations of hydrated IL to the major and minor grooves through hydrophobic and polar interactions resist the DNA to bend. As a result DDd becomes stiffer in higher concentration of IL. For most of the cases stretch modulus of DDd increases with increase of IL solution concentration. This clearly indicates that stiffer DNA requires more force to stretch. Our study shows that both stretch modulus and persistence length changes in the similar fashion with the IL concentration, which is different as observed in case of monovalent salt concentration \cite{baumann97,wu15,garai15}. This trend suggests that DDd behaves more like an elastic rigid rod \cite{kratky49, garai15} in IL. Such a unique behavior of DNA in IL/water binary mixtures can have important implications on DNA stability. Earlier studies by our group showed that dehydration of DNA by ILs indirectly aids in DNA stability against DNase I-mediated hydrolytic damage \cite{senapati12}. The current study suggests that ILs could further aid in stabilizing DNA by directly interfering with DNA-DNase I interaction. It is well known that flexibility of the DNA substrate is key in mediating appropriate DNase I binding and cleavage \cite{heddi10}. Thus, increased rigidity of DNA in ILs might prevent DNase I binding to DNA, thereby conferring exceptional long-term stability of DNA during storage. In principle, outcomes of our study can be tested by performing {\it in vitro} single molecule experiments and MD simulations of DNA-protein interactions in presence of ILs. Our study further provides more insight about DNA nano-mechanics, DNA nanotechnology and IL-DNA interactions, which may help to choose specific nucleic acid solutes for long term DNA storage e.g. {\it [bmim][nitrate]}.

\section{Supplementary Material}
See supplementary material for chemical structures of ILs, a table that describes details of DDd simulated in different IL/water binary mixtures, figures related to time evolution of simulated DDd in $80:20$ and $50:50$ IL/water binary mixtures, figure of the variation of persistence length of DDd at different ILs concentrations, and figure of the variation of stretch modulus of DDd at different ILs concentrations. 

\begin{acknowledgments}
We greatly acknowledge the financial support received from the DAE, India. AG also thanks SERB, DST, government of India, for support through an ECR grant.
\end{acknowledgments}

\nocite{*}
\bibliography{ggsm}

\begin{thebibliography}{59}
\expandafter\ifx\csname natexlab\endcsname\relax\def\natexlab#1{#1}\fi
\expandafter\ifx\csname bibnamefont\endcsname\relax
  \def\bibnamefont#1{#1}\fi
\expandafter\ifx\csname bibfnamefont\endcsname\relax
  \def\bibfnamefont#1{#1}\fi
\expandafter\ifx\csname citenamefont\endcsname\relax
  \def\citenamefont#1{#1}\fi
\expandafter\ifx\csname url\endcsname\relax
  \def\url#1{\texttt{#1}}\fi
\expandafter\ifx\csname urlprefix\endcsname\relax\def\urlprefix{URL }\fi
\providecommand{\bibinfo}[2]{#2}
\providecommand{\eprint}[2][]{\url{#2}}

\bibitem[{\citenamefont{Alberts et~al.}(2008)\citenamefont{Alberts, Alexander,
  Julian, Martin, Keith, and Peter}}]{alberts08}
\bibinfo{author}{\bibfnamefont{B.}~\bibnamefont{Alberts}},
  \bibinfo{author}{\bibfnamefont{J.}~\bibnamefont{Alexander}},
  \bibinfo{author}{\bibfnamefont{L.}~\bibnamefont{Julian}},
  \bibinfo{author}{\bibfnamefont{R.}~\bibnamefont{Martin}},
  \bibinfo{author}{\bibfnamefont{R.}~\bibnamefont{Keith}}, \bibnamefont{and}
  \bibinfo{author}{\bibfnamefont{W.}~\bibnamefont{Peter}},
  \emph{\bibinfo{title}{Molecular biology of the cell}}
  (\bibinfo{publisher}{Garland science}, \bibinfo{year}{2008}),
  \bibinfo{edition}{5th} ed.

\bibitem[{\citenamefont{Mathew-Fenn et~al.}(2008)\citenamefont{Mathew-Fenn,
  Das, and Harbury}}]{fenn08}
\bibinfo{author}{\bibfnamefont{R.~S.} \bibnamefont{Mathew-Fenn}},
  \bibinfo{author}{\bibfnamefont{R.}~\bibnamefont{Das}}, \bibnamefont{and}
  \bibinfo{author}{\bibfnamefont{P.~A.} \bibnamefont{Harbury}},
  \bibinfo{journal}{Science} \textbf{\bibinfo{volume}{322}},
  \bibinfo{pages}{446} (\bibinfo{year}{2008}).

\bibitem[{\citenamefont{Yuan et~al.}(2008)\citenamefont{Yuan, Chen, Lou, and
  Archer}}]{yuan08}
\bibinfo{author}{\bibfnamefont{C.}~\bibnamefont{Yuan}},
  \bibinfo{author}{\bibfnamefont{H.}~\bibnamefont{Chen}},
  \bibinfo{author}{\bibfnamefont{X.~W.} \bibnamefont{Lou}}, \bibnamefont{and}
  \bibinfo{author}{\bibfnamefont{L.~A.} \bibnamefont{Archer}},
  \bibinfo{journal}{Physical review letters} \textbf{\bibinfo{volume}{100}},
  \bibinfo{pages}{018102} (\bibinfo{year}{2008}).

\bibitem[{\citenamefont{Bomble and Case}(2008)}]{bomble08}
\bibinfo{author}{\bibfnamefont{Y.~J.} \bibnamefont{Bomble}} \bibnamefont{and}
  \bibinfo{author}{\bibfnamefont{D.~A.} \bibnamefont{Case}},
  \bibinfo{journal}{Biopolymers} \textbf{\bibinfo{volume}{89}},
  \bibinfo{pages}{722} (\bibinfo{year}{2008}).

\bibitem[{\citenamefont{Noy and Golestanian}(2012)}]{noy12}
\bibinfo{author}{\bibfnamefont{A.}~\bibnamefont{Noy}} \bibnamefont{and}
  \bibinfo{author}{\bibfnamefont{R.}~\bibnamefont{Golestanian}},
  \bibinfo{journal}{Physical review letters} \textbf{\bibinfo{volume}{109}},
  \bibinfo{pages}{228101} (\bibinfo{year}{2012}).

\bibitem[{\citenamefont{Mogurampelly et~al.}(2013)\citenamefont{Mogurampelly,
  Nandy, Netz, and Maiti}}]{maiti13}
\bibinfo{author}{\bibfnamefont{S.}~\bibnamefont{Mogurampelly}},
  \bibinfo{author}{\bibfnamefont{B.}~\bibnamefont{Nandy}},
  \bibinfo{author}{\bibfnamefont{R.~R.} \bibnamefont{Netz}}, \bibnamefont{and}
  \bibinfo{author}{\bibfnamefont{P.~K.} \bibnamefont{Maiti}},
  \bibinfo{journal}{The European Physical Journal E}
  \textbf{\bibinfo{volume}{36}}, \bibinfo{pages}{68} (\bibinfo{year}{2013}).

\bibitem[{\citenamefont{Mazur and Maaloum}(2014{\natexlab{a}})}]{mazur14}
\bibinfo{author}{\bibfnamefont{A.~K.} \bibnamefont{Mazur}} \bibnamefont{and}
  \bibinfo{author}{\bibfnamefont{M.}~\bibnamefont{Maaloum}},
  \bibinfo{journal}{Physical review letters} \textbf{\bibinfo{volume}{112}},
  \bibinfo{pages}{068104} (\bibinfo{year}{2014}{\natexlab{a}}).

\bibitem[{\citenamefont{Mazur and Maaloum}(2014{\natexlab{b}})}]{maaloum14}
\bibinfo{author}{\bibfnamefont{A.~K.} \bibnamefont{Mazur}} \bibnamefont{and}
  \bibinfo{author}{\bibfnamefont{M.}~\bibnamefont{Maaloum}},
  \bibinfo{journal}{Nucleic acids research} \textbf{\bibinfo{volume}{42}},
  \bibinfo{pages}{14006} (\bibinfo{year}{2014}{\natexlab{b}}).

\bibitem[{\citenamefont{Wu et~al.}(2015)\citenamefont{Wu, Bao, Zhang, and
  Tan}}]{wu15}
\bibinfo{author}{\bibfnamefont{Y.-Y.} \bibnamefont{Wu}},
  \bibinfo{author}{\bibfnamefont{L.}~\bibnamefont{Bao}},
  \bibinfo{author}{\bibfnamefont{X.}~\bibnamefont{Zhang}}, \bibnamefont{and}
  \bibinfo{author}{\bibfnamefont{Z.-J.} \bibnamefont{Tan}},
  \bibinfo{journal}{The Journal of Chemical Physics}
  \textbf{\bibinfo{volume}{142}}, \bibinfo{pages}{125103}
  (\bibinfo{year}{2015}).

\bibitem[{\citenamefont{Garai et~al.}(2015)\citenamefont{Garai, Saurabh,
  Lansac, and Maiti}}]{garai15}
\bibinfo{author}{\bibfnamefont{A.}~\bibnamefont{Garai}},
  \bibinfo{author}{\bibfnamefont{S.}~\bibnamefont{Saurabh}},
  \bibinfo{author}{\bibfnamefont{Y.}~\bibnamefont{Lansac}}, \bibnamefont{and}
  \bibinfo{author}{\bibfnamefont{P.~K.} \bibnamefont{Maiti}},
  \bibinfo{journal}{The Journal of Physical Chemistry B}
  \textbf{\bibinfo{volume}{119}}, \bibinfo{pages}{11146}
  (\bibinfo{year}{2015}).

\bibitem[{\citenamefont{Cloutier and Widom}(2004)}]{cloutier04}
\bibinfo{author}{\bibfnamefont{T.~E.} \bibnamefont{Cloutier}} \bibnamefont{and}
  \bibinfo{author}{\bibfnamefont{J.}~\bibnamefont{Widom}},
  \bibinfo{journal}{Molecular cell} \textbf{\bibinfo{volume}{14}},
  \bibinfo{pages}{355} (\bibinfo{year}{2004}).

\bibitem[{\citenamefont{Wiggins et~al.}(2006)\citenamefont{Wiggins, Van
  Der~Heijden, Moreno-Herrero, Spakowitz, Phillips, Widom, Dekker, and
  Nelson}}]{wiggins06}
\bibinfo{author}{\bibfnamefont{P.~A.} \bibnamefont{Wiggins}},
  \bibinfo{author}{\bibfnamefont{T.}~\bibnamefont{Van Der~Heijden}},
  \bibinfo{author}{\bibfnamefont{F.}~\bibnamefont{Moreno-Herrero}},
  \bibinfo{author}{\bibfnamefont{A.}~\bibnamefont{Spakowitz}},
  \bibinfo{author}{\bibfnamefont{R.}~\bibnamefont{Phillips}},
  \bibinfo{author}{\bibfnamefont{J.}~\bibnamefont{Widom}},
  \bibinfo{author}{\bibfnamefont{C.}~\bibnamefont{Dekker}}, \bibnamefont{and}
  \bibinfo{author}{\bibfnamefont{P.~C.} \bibnamefont{Nelson}},
  \bibinfo{journal}{Nature nanotechnology} \textbf{\bibinfo{volume}{1}},
  \bibinfo{pages}{137} (\bibinfo{year}{2006}).

\bibitem[{\citenamefont{Vafabakhsh and Ha}(2012)}]{vafabakhsh12}
\bibinfo{author}{\bibfnamefont{R.}~\bibnamefont{Vafabakhsh}} \bibnamefont{and}
  \bibinfo{author}{\bibfnamefont{T.}~\bibnamefont{Ha}},
  \bibinfo{journal}{Science} \textbf{\bibinfo{volume}{337}},
  \bibinfo{pages}{1097} (\bibinfo{year}{2012}).

\bibitem[{\citenamefont{Baumann et~al.}(1997)\citenamefont{Baumann, Smith,
  Bloomfield, and Bustamante}}]{baumann97}
\bibinfo{author}{\bibfnamefont{C.~G.} \bibnamefont{Baumann}},
  \bibinfo{author}{\bibfnamefont{S.~B.} \bibnamefont{Smith}},
  \bibinfo{author}{\bibfnamefont{V.~A.} \bibnamefont{Bloomfield}},
  \bibnamefont{and}
  \bibinfo{author}{\bibfnamefont{C.}~\bibnamefont{Bustamante}},
  \bibinfo{journal}{Proceedings of the National Academy of Sciences}
  \textbf{\bibinfo{volume}{94}}, \bibinfo{pages}{6185} (\bibinfo{year}{1997}).

\bibitem[{\citenamefont{Marko and Siggia}(1995)}]{marko95}
\bibinfo{author}{\bibfnamefont{J.~F.} \bibnamefont{Marko}} \bibnamefont{and}
  \bibinfo{author}{\bibfnamefont{E.~D.} \bibnamefont{Siggia}},
  \bibinfo{journal}{Macromolecules} \textbf{\bibinfo{volume}{28}},
  \bibinfo{pages}{8759} (\bibinfo{year}{1995}).

\bibitem[{\citenamefont{Kumar and Li}(2010)}]{kumar10}
\bibinfo{author}{\bibfnamefont{S.}~\bibnamefont{Kumar}} \bibnamefont{and}
  \bibinfo{author}{\bibfnamefont{M.~S.} \bibnamefont{Li}},
  \bibinfo{journal}{Physics Reports} \textbf{\bibinfo{volume}{486}},
  \bibinfo{pages}{1} (\bibinfo{year}{2010}).

\bibitem[{\citenamefont{Patil et~al.}(2005)\citenamefont{Patil, Rhodes, and
  Burgess}}]{patil05}
\bibinfo{author}{\bibfnamefont{S.~D.} \bibnamefont{Patil}},
  \bibinfo{author}{\bibfnamefont{D.~G.} \bibnamefont{Rhodes}},
  \bibnamefont{and} \bibinfo{author}{\bibfnamefont{D.~J.}
  \bibnamefont{Burgess}}, \bibinfo{journal}{The AAPS journal}
  \textbf{\bibinfo{volume}{7}}, \bibinfo{pages}{E61} (\bibinfo{year}{2005}).

\bibitem[{\citenamefont{Liu}(2003)}]{liu03}
\bibinfo{author}{\bibfnamefont{M.}~\bibnamefont{Liu}},
  \bibinfo{journal}{Journal of internal medicine}
  \textbf{\bibinfo{volume}{253}}, \bibinfo{pages}{402} (\bibinfo{year}{2003}).

\bibitem[{\citenamefont{Goldman et~al.}(2013)\citenamefont{Goldman, Bertone,
  Chen, Dessimoz, LeProust, Sipos, and Birney}}]{goldman13}
\bibinfo{author}{\bibfnamefont{N.}~\bibnamefont{Goldman}},
  \bibinfo{author}{\bibfnamefont{P.}~\bibnamefont{Bertone}},
  \bibinfo{author}{\bibfnamefont{S.}~\bibnamefont{Chen}},
  \bibinfo{author}{\bibfnamefont{C.}~\bibnamefont{Dessimoz}},
  \bibinfo{author}{\bibfnamefont{E.~M.} \bibnamefont{LeProust}},
  \bibinfo{author}{\bibfnamefont{B.}~\bibnamefont{Sipos}}, \bibnamefont{and}
  \bibinfo{author}{\bibfnamefont{E.}~\bibnamefont{Birney}},
  \bibinfo{journal}{Nature} \textbf{\bibinfo{volume}{494}}, \bibinfo{pages}{77}
  (\bibinfo{year}{2013}).

\bibitem[{\citenamefont{Normile}(2002)}]{normile02}
\bibinfo{author}{\bibfnamefont{D.}~\bibnamefont{Normile}},
  \bibinfo{journal}{Science} \textbf{\bibinfo{volume}{295}},
  \bibinfo{pages}{951} (\bibinfo{year}{2002}).

\bibitem[{\citenamefont{Joshi et~al.}(2016)\citenamefont{Joshi, Kaushik,
  Seeman, and Maiti}}]{maiti16}
\bibinfo{author}{\bibfnamefont{H.}~\bibnamefont{Joshi}},
  \bibinfo{author}{\bibfnamefont{A.}~\bibnamefont{Kaushik}},
  \bibinfo{author}{\bibfnamefont{N.~C.} \bibnamefont{Seeman}},
  \bibnamefont{and} \bibinfo{author}{\bibfnamefont{P.~K.} \bibnamefont{Maiti}},
  \bibinfo{journal}{ACS nano} \textbf{\bibinfo{volume}{10}},
  \bibinfo{pages}{7780} (\bibinfo{year}{2016}).

\bibitem[{\citenamefont{Joshi et~al.}(2017)\citenamefont{Joshi, Bhatia,
  Krishnan, and Maiti}}]{maiti17}
\bibinfo{author}{\bibfnamefont{H.}~\bibnamefont{Joshi}},
  \bibinfo{author}{\bibfnamefont{D.}~\bibnamefont{Bhatia}},
  \bibinfo{author}{\bibfnamefont{Y.}~\bibnamefont{Krishnan}}, \bibnamefont{and}
  \bibinfo{author}{\bibfnamefont{P.~K.} \bibnamefont{Maiti}},
  \bibinfo{journal}{Nanoscale} \textbf{\bibinfo{volume}{9}},
  \bibinfo{pages}{4467} (\bibinfo{year}{2017}).

\bibitem[{\citenamefont{Joshi et~al.}(2015)\citenamefont{Joshi, Dwaraknath, and
  Maiti}}]{maiti15}
\bibinfo{author}{\bibfnamefont{H.}~\bibnamefont{Joshi}},
  \bibinfo{author}{\bibfnamefont{A.}~\bibnamefont{Dwaraknath}},
  \bibnamefont{and} \bibinfo{author}{\bibfnamefont{P.~K.} \bibnamefont{Maiti}},
  \bibinfo{journal}{Physical Chemistry Chemical Physics}
  \textbf{\bibinfo{volume}{17}}, \bibinfo{pages}{1424} (\bibinfo{year}{2015}).

\bibitem[{\citenamefont{Maiti et~al.}(2006)\citenamefont{Maiti, Pascal,
  Vaidehi, Heo, and Goddard~III}}]{maiti06}
\bibinfo{author}{\bibfnamefont{P.~K.} \bibnamefont{Maiti}},
  \bibinfo{author}{\bibfnamefont{T.~A.} \bibnamefont{Pascal}},
  \bibinfo{author}{\bibfnamefont{N.}~\bibnamefont{Vaidehi}},
  \bibinfo{author}{\bibfnamefont{J.}~\bibnamefont{Heo}}, \bibnamefont{and}
  \bibinfo{author}{\bibfnamefont{W.~A.} \bibnamefont{Goddard~III}},
  \bibinfo{journal}{Biophysical journal} \textbf{\bibinfo{volume}{90}},
  \bibinfo{pages}{1463} (\bibinfo{year}{2006}).

\bibitem[{\citenamefont{Maiti et~al.}(2004)\citenamefont{Maiti, Pascal,
  Vaidehi, and Goddard~III}}]{maiti04}
\bibinfo{author}{\bibfnamefont{P.~K.} \bibnamefont{Maiti}},
  \bibinfo{author}{\bibfnamefont{T.~A.} \bibnamefont{Pascal}},
  \bibinfo{author}{\bibfnamefont{N.}~\bibnamefont{Vaidehi}}, \bibnamefont{and}
  \bibinfo{author}{\bibfnamefont{W.~A.} \bibnamefont{Goddard~III}},
  \bibinfo{journal}{Nucleic acids research} \textbf{\bibinfo{volume}{32}},
  \bibinfo{pages}{6047} (\bibinfo{year}{2004}).

\bibitem[{\citenamefont{Liu et~al.}(2004)\citenamefont{Liu, Wang, Dou, Wang,
  Xie, Yin, Zhang, and Xi}}]{liu04}
\bibinfo{author}{\bibfnamefont{Y.-Y.} \bibnamefont{Liu}},
  \bibinfo{author}{\bibfnamefont{P.-Y.} \bibnamefont{Wang}},
  \bibinfo{author}{\bibfnamefont{S.-X.} \bibnamefont{Dou}},
  \bibinfo{author}{\bibfnamefont{W.-C.} \bibnamefont{Wang}},
  \bibinfo{author}{\bibfnamefont{P.}~\bibnamefont{Xie}},
  \bibinfo{author}{\bibfnamefont{H.-W.} \bibnamefont{Yin}},
  \bibinfo{author}{\bibfnamefont{X.-D.} \bibnamefont{Zhang}}, \bibnamefont{and}
  \bibinfo{author}{\bibfnamefont{X.~G.} \bibnamefont{Xi}},
  \bibinfo{journal}{The Journal of chemical physics}
  \textbf{\bibinfo{volume}{121}}, \bibinfo{pages}{4302} (\bibinfo{year}{2004}).

\bibitem[{\citenamefont{Liu and Liu}(2009)}]{liu09}
\bibinfo{author}{\bibfnamefont{H.}~\bibnamefont{Liu}} \bibnamefont{and}
  \bibinfo{author}{\bibfnamefont{D.}~\bibnamefont{Liu}},
  \bibinfo{journal}{Chemical Communications} pp. \bibinfo{pages}{2625--2636}
  (\bibinfo{year}{2009}).

\bibitem[{\citenamefont{Yamada et~al.}(2005)\citenamefont{Yamada, Yokota, Kaya,
  Satoh, Jonganurakkun, Nomizu, and Nishi}}]{yamada05}
\bibinfo{author}{\bibfnamefont{M.}~\bibnamefont{Yamada}},
  \bibinfo{author}{\bibfnamefont{M.}~\bibnamefont{Yokota}},
  \bibinfo{author}{\bibfnamefont{M.}~\bibnamefont{Kaya}},
  \bibinfo{author}{\bibfnamefont{S.}~\bibnamefont{Satoh}},
  \bibinfo{author}{\bibfnamefont{B.}~\bibnamefont{Jonganurakkun}},
  \bibinfo{author}{\bibfnamefont{M.}~\bibnamefont{Nomizu}}, \bibnamefont{and}
  \bibinfo{author}{\bibfnamefont{N.}~\bibnamefont{Nishi}},
  \bibinfo{journal}{Polymer} \textbf{\bibinfo{volume}{46}},
  \bibinfo{pages}{10102} (\bibinfo{year}{2005}).

\bibitem[{\citenamefont{Cheng and Pettitt}(1992)}]{cheng92}
\bibinfo{author}{\bibfnamefont{Y.-K.} \bibnamefont{Cheng}} \bibnamefont{and}
  \bibinfo{author}{\bibfnamefont{B.~M.} \bibnamefont{Pettitt}},
  \bibinfo{journal}{Progress in biophysics and molecular biology}
  \textbf{\bibinfo{volume}{58}}, \bibinfo{pages}{225} (\bibinfo{year}{1992}).

\bibitem[{\citenamefont{Lindahl and Nyberg}(1972)}]{lindahl72}
\bibinfo{author}{\bibfnamefont{T.}~\bibnamefont{Lindahl}} \bibnamefont{and}
  \bibinfo{author}{\bibfnamefont{B.}~\bibnamefont{Nyberg}},
  \bibinfo{journal}{Biochemistry} \textbf{\bibinfo{volume}{11}},
  \bibinfo{pages}{3610} (\bibinfo{year}{1972}).

\bibitem[{\citenamefont{Tateishi-Karimata and Sugimoto}(2014)}]{sugimoto14}
\bibinfo{author}{\bibfnamefont{H.}~\bibnamefont{Tateishi-Karimata}}
  \bibnamefont{and} \bibinfo{author}{\bibfnamefont{N.}~\bibnamefont{Sugimoto}},
  \bibinfo{journal}{Nucleic acids research} \textbf{\bibinfo{volume}{42}},
  \bibinfo{pages}{8831} (\bibinfo{year}{2014}).

\bibitem[{\citenamefont{Bonner and Klibanov}(2000)}]{bonner00}
\bibinfo{author}{\bibfnamefont{G.}~\bibnamefont{Bonner}} \bibnamefont{and}
  \bibinfo{author}{\bibfnamefont{A.~M.} \bibnamefont{Klibanov}},
  \bibinfo{journal}{Biotechnology and bioengineering}
  \textbf{\bibinfo{volume}{68}}, \bibinfo{pages}{339} (\bibinfo{year}{2000}).

\bibitem[{\citenamefont{Hammouda and Worcester}(2006)}]{hammouda06}
\bibinfo{author}{\bibfnamefont{B.}~\bibnamefont{Hammouda}} \bibnamefont{and}
  \bibinfo{author}{\bibfnamefont{D.}~\bibnamefont{Worcester}},
  \bibinfo{journal}{Biophysical journal} \textbf{\bibinfo{volume}{91}},
  \bibinfo{pages}{2237} (\bibinfo{year}{2006}).

\bibitem[{\citenamefont{Vijayaraghavan
  et~al.}(2010)\citenamefont{Vijayaraghavan, Izgorodin, Ganesh, Surianarayanan,
  and MacFarlane}}]{macfalrane10}
\bibinfo{author}{\bibfnamefont{R.}~\bibnamefont{Vijayaraghavan}},
  \bibinfo{author}{\bibfnamefont{A.}~\bibnamefont{Izgorodin}},
  \bibinfo{author}{\bibfnamefont{V.}~\bibnamefont{Ganesh}},
  \bibinfo{author}{\bibfnamefont{M.}~\bibnamefont{Surianarayanan}},
  \bibnamefont{and} \bibinfo{author}{\bibfnamefont{D.~R.}
  \bibnamefont{MacFarlane}}, \bibinfo{journal}{Angewandte Chemie International
  Edition} \textbf{\bibinfo{volume}{49}}, \bibinfo{pages}{1631}
  (\bibinfo{year}{2010}).

\bibitem[{\citenamefont{Chandran et~al.}(2012)\citenamefont{Chandran,
  Ghoshdastidar, and Senapati}}]{senapati12}
\bibinfo{author}{\bibfnamefont{A.}~\bibnamefont{Chandran}},
  \bibinfo{author}{\bibfnamefont{D.}~\bibnamefont{Ghoshdastidar}},
  \bibnamefont{and} \bibinfo{author}{\bibfnamefont{S.}~\bibnamefont{Senapati}},
  \bibinfo{journal}{Journal of the American Chemical Society}
  \textbf{\bibinfo{volume}{134}}, \bibinfo{pages}{20330}
  (\bibinfo{year}{2012}).

\bibitem[{\citenamefont{Liu et~al.}(2012)\citenamefont{Liu, Maginn, Visser,
  Bridges, and Fox}}]{liu12}
\bibinfo{author}{\bibfnamefont{H.}~\bibnamefont{Liu}},
  \bibinfo{author}{\bibfnamefont{E.}~\bibnamefont{Maginn}},
  \bibinfo{author}{\bibfnamefont{A.~E.} \bibnamefont{Visser}},
  \bibinfo{author}{\bibfnamefont{N.~J.} \bibnamefont{Bridges}},
  \bibnamefont{and} \bibinfo{author}{\bibfnamefont{E.~B.} \bibnamefont{Fox}},
  \bibinfo{journal}{Industrial \& Engineering Chemistry Research}
  \textbf{\bibinfo{volume}{51}}, \bibinfo{pages}{7242} (\bibinfo{year}{2012}).

\bibitem[{\citenamefont{Hayes et~al.}(2015)\citenamefont{Hayes, Warr, and
  Atkin}}]{hayes15}
\bibinfo{author}{\bibfnamefont{R.}~\bibnamefont{Hayes}},
  \bibinfo{author}{\bibfnamefont{G.~G.} \bibnamefont{Warr}}, \bibnamefont{and}
  \bibinfo{author}{\bibfnamefont{R.}~\bibnamefont{Atkin}},
  \bibinfo{journal}{Chemical reviews} \textbf{\bibinfo{volume}{115}},
  \bibinfo{pages}{6357} (\bibinfo{year}{2015}).

\bibitem[{\citenamefont{Lavery et~al.}(2009)\citenamefont{Lavery, Moakher,
  Maddocks, Petkeviciute, and Zakrzewska}}]{lavery09}
\bibinfo{author}{\bibfnamefont{R.}~\bibnamefont{Lavery}},
  \bibinfo{author}{\bibfnamefont{M.}~\bibnamefont{Moakher}},
  \bibinfo{author}{\bibfnamefont{J.~H.} \bibnamefont{Maddocks}},
  \bibinfo{author}{\bibfnamefont{D.}~\bibnamefont{Petkeviciute}},
  \bibnamefont{and}
  \bibinfo{author}{\bibfnamefont{K.}~\bibnamefont{Zakrzewska}},
  \bibinfo{journal}{Nucleic acids research} \textbf{\bibinfo{volume}{37}},
  \bibinfo{pages}{5917} (\bibinfo{year}{2009}).

\bibitem[{\citenamefont{P{\'e}rez
  et~al.}(2007{\natexlab{a}})\citenamefont{P{\'e}rez, Luque, and
  Orozco}}]{perez07}
\bibinfo{author}{\bibfnamefont{A.}~\bibnamefont{P{\'e}rez}},
  \bibinfo{author}{\bibfnamefont{F.~J.} \bibnamefont{Luque}}, \bibnamefont{and}
  \bibinfo{author}{\bibfnamefont{M.}~\bibnamefont{Orozco}},
  \bibinfo{journal}{Journal of the American Chemical Society}
  \textbf{\bibinfo{volume}{129}}, \bibinfo{pages}{14739}
  (\bibinfo{year}{2007}{\natexlab{a}}).

\bibitem[{\citenamefont{Cheatham~III et~al.}(1999)\citenamefont{Cheatham~III,
  Cieplak, and Kollman}}]{kollman99}
\bibinfo{author}{\bibfnamefont{T.~E.} \bibnamefont{Cheatham~III}},
  \bibinfo{author}{\bibfnamefont{P.}~\bibnamefont{Cieplak}}, \bibnamefont{and}
  \bibinfo{author}{\bibfnamefont{P.~A.} \bibnamefont{Kollman}},
  \bibinfo{journal}{Journal of Biomolecular Structure and Dynamics}
  \textbf{\bibinfo{volume}{16}}, \bibinfo{pages}{845} (\bibinfo{year}{1999}).

\bibitem[{\citenamefont{P{\'e}rez
  et~al.}(2007{\natexlab{b}})\citenamefont{P{\'e}rez, March{\'a}n, Svozil,
  Sponer, Cheatham~III, Laughton, and Orozco}}]{orozco07}
\bibinfo{author}{\bibfnamefont{A.}~\bibnamefont{P{\'e}rez}},
  \bibinfo{author}{\bibfnamefont{I.}~\bibnamefont{March{\'a}n}},
  \bibinfo{author}{\bibfnamefont{D.}~\bibnamefont{Svozil}},
  \bibinfo{author}{\bibfnamefont{J.}~\bibnamefont{Sponer}},
  \bibinfo{author}{\bibfnamefont{T.~E.} \bibnamefont{Cheatham~III}},
  \bibinfo{author}{\bibfnamefont{C.~A.} \bibnamefont{Laughton}},
  \bibnamefont{and} \bibinfo{author}{\bibfnamefont{M.}~\bibnamefont{Orozco}},
  \bibinfo{journal}{Biophysical journal} \textbf{\bibinfo{volume}{92}},
  \bibinfo{pages}{3817} (\bibinfo{year}{2007}{\natexlab{b}}).

\bibitem[{\citenamefont{Jorgensen et~al.}(1983)\citenamefont{Jorgensen,
  Chandrasekhar, Madura, Impey, and Klein}}]{klein83}
\bibinfo{author}{\bibfnamefont{W.~L.} \bibnamefont{Jorgensen}},
  \bibinfo{author}{\bibfnamefont{J.}~\bibnamefont{Chandrasekhar}},
  \bibinfo{author}{\bibfnamefont{J.~D.} \bibnamefont{Madura}},
  \bibinfo{author}{\bibfnamefont{R.~W.} \bibnamefont{Impey}}, \bibnamefont{and}
  \bibinfo{author}{\bibfnamefont{M.~L.} \bibnamefont{Klein}},
  \bibinfo{journal}{The Journal of chemical physics}
  \textbf{\bibinfo{volume}{79}}, \bibinfo{pages}{926} (\bibinfo{year}{1983}).

\bibitem[{\citenamefont{Canongia~Lopes and Padua}(2006)}]{lopes06}
\bibinfo{author}{\bibfnamefont{J.~N.} \bibnamefont{Canongia~Lopes}}
  \bibnamefont{and} \bibinfo{author}{\bibfnamefont{A.~A.} \bibnamefont{Padua}},
  \bibinfo{journal}{The Journal of Physical Chemistry B}
  \textbf{\bibinfo{volume}{110}}, \bibinfo{pages}{3330} (\bibinfo{year}{2006}).

\bibitem[{\citenamefont{Sambasivarao and Acevedo}(2009)}]{somisetti09}
\bibinfo{author}{\bibfnamefont{S.~V.} \bibnamefont{Sambasivarao}}
  \bibnamefont{and} \bibinfo{author}{\bibfnamefont{O.}~\bibnamefont{Acevedo}},
  \bibinfo{journal}{Journal of chemical theory and computation}
  \textbf{\bibinfo{volume}{5}}, \bibinfo{pages}{1038} (\bibinfo{year}{2009}).

\bibitem[{\citenamefont{Cheatham and Case}(2013)}]{cheatham13}
\bibinfo{author}{\bibfnamefont{T.~E.} \bibnamefont{Cheatham}} \bibnamefont{and}
  \bibinfo{author}{\bibfnamefont{D.~A.} \bibnamefont{Case}},
  \bibinfo{journal}{Biopolymers} \textbf{\bibinfo{volume}{99}},
  \bibinfo{pages}{969} (\bibinfo{year}{2013}).

\bibitem[{\citenamefont{Kratky and Porod}(1949)}]{kratky49}
\bibinfo{author}{\bibfnamefont{O.}~\bibnamefont{Kratky}} \bibnamefont{and}
  \bibinfo{author}{\bibfnamefont{G.}~\bibnamefont{Porod}},
  \bibinfo{journal}{Journal of colloid science} \textbf{\bibinfo{volume}{4}},
  \bibinfo{pages}{35} (\bibinfo{year}{1949}).

\bibitem[{\citenamefont{Zoli}(2018)}]{zoli18}
\bibinfo{author}{\bibfnamefont{M.}~\bibnamefont{Zoli}}, \bibinfo{journal}{The
  Journal of chemical physics} \textbf{\bibinfo{volume}{148}},
  \bibinfo{pages}{214902} (\bibinfo{year}{2018}).

\bibitem[{\citenamefont{Zoli}(2016{\natexlab{a}})}]{zoli16}
\bibinfo{author}{\bibfnamefont{M.}~\bibnamefont{Zoli}},
  \bibinfo{journal}{Physical Chemistry Chemical Physics}
  \textbf{\bibinfo{volume}{18}}, \bibinfo{pages}{17666}
  (\bibinfo{year}{2016}{\natexlab{a}}).

\bibitem[{\citenamefont{Zoli}(2016{\natexlab{b}})}]{zoli2016}
\bibinfo{author}{\bibfnamefont{M.}~\bibnamefont{Zoli}}, \bibinfo{journal}{The
  Journal of Chemical Physics} \textbf{\bibinfo{volume}{144}},
  \bibinfo{pages}{214104} (\bibinfo{year}{2016}{\natexlab{b}}).

\bibitem[{\citenamefont{Mazur}(2007)}]{mazur07}
\bibinfo{author}{\bibfnamefont{A.~K.} \bibnamefont{Mazur}},
  \bibinfo{journal}{Physical review letters} \textbf{\bibinfo{volume}{98}},
  \bibinfo{pages}{218102} (\bibinfo{year}{2007}).

\bibitem[{\citenamefont{Mazur and Maaloum}(2014{\natexlab{c}})}]{mazur2014}
\bibinfo{author}{\bibfnamefont{A.~K.} \bibnamefont{Mazur}} \bibnamefont{and}
  \bibinfo{author}{\bibfnamefont{M.}~\bibnamefont{Maaloum}},
  \bibinfo{journal}{Physical review letters} \textbf{\bibinfo{volume}{112}},
  \bibinfo{pages}{068104} (\bibinfo{year}{2014}{\natexlab{c}}).

\bibitem[{\citenamefont{Marko}(2015)}]{marko15}
\bibinfo{author}{\bibfnamefont{J.~F.} \bibnamefont{Marko}},
  \bibinfo{journal}{Physica A: Statistical Mechanics and its Applications}
  \textbf{\bibinfo{volume}{418}}, \bibinfo{pages}{126} (\bibinfo{year}{2015}).

\bibitem[{\citenamefont{Bustamante et~al.}(1994)\citenamefont{Bustamante,
  Marko, Siggia, and Smith}}]{bustamante94}
\bibinfo{author}{\bibfnamefont{C.}~\bibnamefont{Bustamante}},
  \bibinfo{author}{\bibfnamefont{J.~F.} \bibnamefont{Marko}},
  \bibinfo{author}{\bibfnamefont{E.~D.} \bibnamefont{Siggia}},
  \bibnamefont{and} \bibinfo{author}{\bibfnamefont{S.}~\bibnamefont{Smith}},
  \bibinfo{journal}{Science- New York Then Washington-}
  \textbf{\bibinfo{volume}{265}}, \bibinfo{pages}{1599} (\bibinfo{year}{1994}).

\bibitem[{\citenamefont{Galindo-Murillo
  et~al.}(2014)\citenamefont{Galindo-Murillo, Roe, and
  Cheatham~III}}]{galindo14}
\bibinfo{author}{\bibfnamefont{R.}~\bibnamefont{Galindo-Murillo}},
  \bibinfo{author}{\bibfnamefont{D.~R.} \bibnamefont{Roe}}, \bibnamefont{and}
  \bibinfo{author}{\bibfnamefont{T.~E.} \bibnamefont{Cheatham~III}},
  \bibinfo{journal}{Nature communications} \textbf{\bibinfo{volume}{5}},
  \bibinfo{pages}{5152} (\bibinfo{year}{2014}).

\bibitem[{\citenamefont{Singh et~al.}(2017)\citenamefont{Singh, Gardas, and
  Senapati}}]{senapati17}
\bibinfo{author}{\bibfnamefont{A.~P.} \bibnamefont{Singh}},
  \bibinfo{author}{\bibfnamefont{R.~L.} \bibnamefont{Gardas}},
  \bibnamefont{and} \bibinfo{author}{\bibfnamefont{S.}~\bibnamefont{Senapati}},
  \bibinfo{journal}{Soft matter} \textbf{\bibinfo{volume}{13}},
  \bibinfo{pages}{2348} (\bibinfo{year}{2017}).

\bibitem[{\citenamefont{Udachin and Ripmeester}(1999)}]{konstantin99}
\bibinfo{author}{\bibfnamefont{K.~A.} \bibnamefont{Udachin}} \bibnamefont{and}
  \bibinfo{author}{\bibfnamefont{J.~A.} \bibnamefont{Ripmeester}},
  \bibinfo{journal}{Nature} \textbf{\bibinfo{volume}{397}},
  \bibinfo{pages}{420} (\bibinfo{year}{1999}).

\bibitem[{\citenamefont{Dom{\'\i}nguez
  et~al.}(2017)\citenamefont{Dom{\'\i}nguez, Ramos, Mendieta-Moreno, Fierro,
  Mendieta, Tamayo, and Calleja}}]{carmen17}
\bibinfo{author}{\bibfnamefont{C.~M.} \bibnamefont{Dom{\'\i}nguez}},
  \bibinfo{author}{\bibfnamefont{D.}~\bibnamefont{Ramos}},
  \bibinfo{author}{\bibfnamefont{J.~I.} \bibnamefont{Mendieta-Moreno}},
  \bibinfo{author}{\bibfnamefont{J.~L.} \bibnamefont{Fierro}},
  \bibinfo{author}{\bibfnamefont{J.}~\bibnamefont{Mendieta}},
  \bibinfo{author}{\bibfnamefont{J.}~\bibnamefont{Tamayo}}, \bibnamefont{and}
  \bibinfo{author}{\bibfnamefont{M.}~\bibnamefont{Calleja}},
  \bibinfo{journal}{Scientific Reports} \textbf{\bibinfo{volume}{7}},
  \bibinfo{pages}{536} (\bibinfo{year}{2017}).

\bibitem[{\citenamefont{Lanka{\v{s}} et~al.}(2003)\citenamefont{Lanka{\v{s}},
  {\v{S}}poner, Langowski, and Cheatham~III}}]{lankavs03}
\bibinfo{author}{\bibfnamefont{F.}~\bibnamefont{Lanka{\v{s}}}},
  \bibinfo{author}{\bibfnamefont{J.}~\bibnamefont{{\v{S}}poner}},
  \bibinfo{author}{\bibfnamefont{J.}~\bibnamefont{Langowski}},
  \bibnamefont{and} \bibinfo{author}{\bibfnamefont{T.~E.}
  \bibnamefont{Cheatham~III}}, \bibinfo{journal}{Biophysical journal}
  \textbf{\bibinfo{volume}{85}}, \bibinfo{pages}{2872} (\bibinfo{year}{2003}).

\bibitem[{\citenamefont{Heddi et~al.}(2010)\citenamefont{Heddi, Abi-Ghanem,
  Lavigne, and Hartmann}}]{heddi10}
\bibinfo{author}{\bibfnamefont{B.}~\bibnamefont{Heddi}},
  \bibinfo{author}{\bibfnamefont{J.}~\bibnamefont{Abi-Ghanem}},
  \bibinfo{author}{\bibfnamefont{M.}~\bibnamefont{Lavigne}}, \bibnamefont{and}
  \bibinfo{author}{\bibfnamefont{B.}~\bibnamefont{Hartmann}},
  \bibinfo{journal}{Journal of molecular biology}
  \textbf{\bibinfo{volume}{395}}, \bibinfo{pages}{123} (\bibinfo{year}{2010}).

\end{thebibliography}

\end{document}


\title{\hspace{3cm} Supplementary Material for \newline \hfill \break ``Ionic liquids make DNA rigid"}
\author{Ashok Garai}
\address{Centre for Condensed Matter Theory, Department of Physics, Indian Institute of Science, Bangalore 560012, India}
\affiliation{Department of Physics, LNM Institute of Information Technology, Jamdoli, Jaipur 302031, India}
\author{Debostuti Ghoshdastidar}
\address{Bhupat and Jyoti Mehta School of Biosciences, Department of Biotechnology, Indian Institute of Technology Madras, Chennai 600036, India}
\author{Sanjib Senapati}
\address{Bhupat and Jyoti Mehta School of Biosciences, Department of Biotechnology, Indian Institute of Technology Madras, Chennai 600036, India}
\author{Prabal K Maiti}
\affiliation{Centre for Condensed Matter Theory, Department of Physics, Indian Institute of Science, Bangalore 560012, India}
\date{\today}

\maketitle

\begin{figure}
\centering
\includegraphics{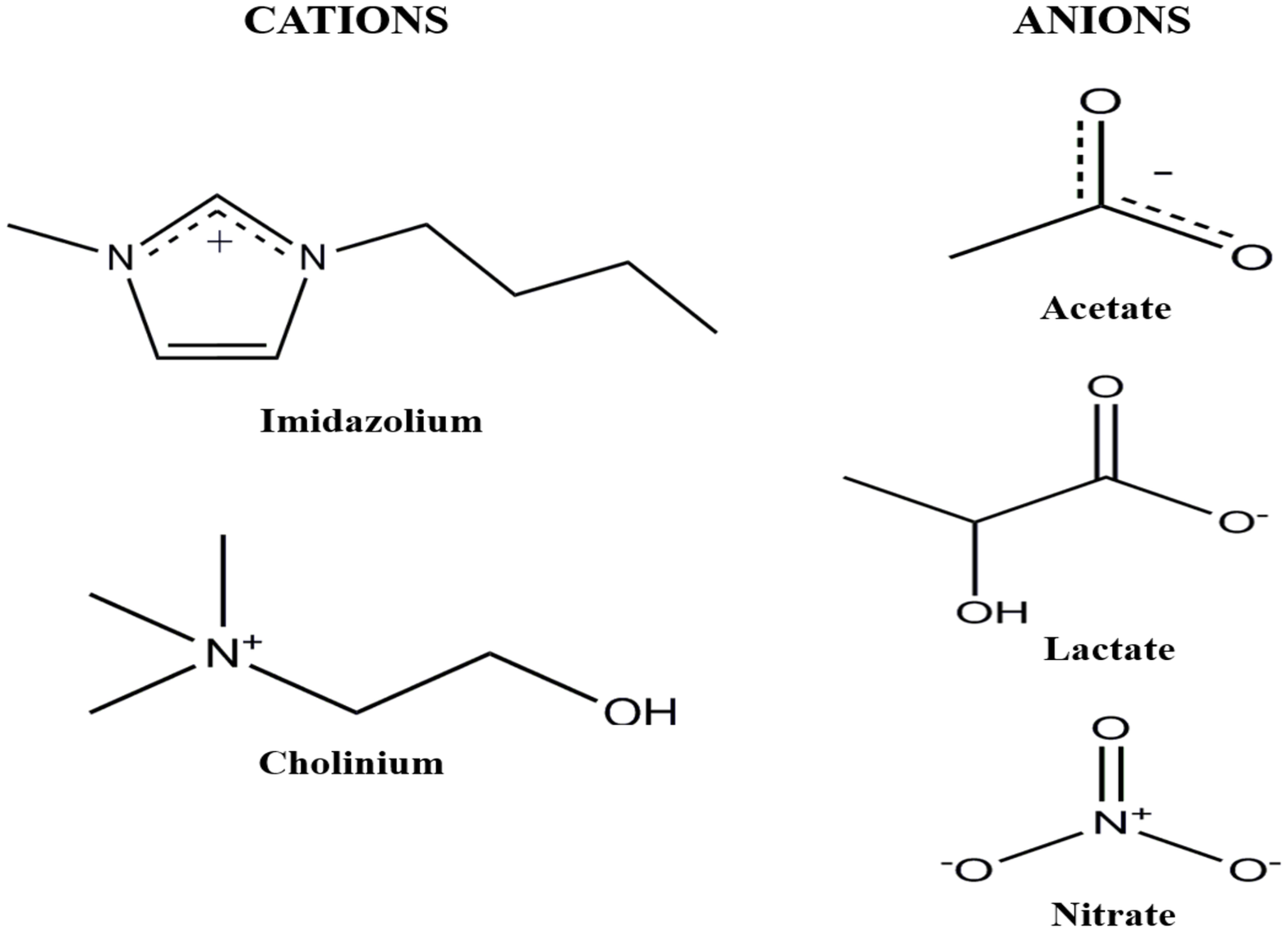}
\caption{Chemical structures of IL cations and anions used in this study. The positively charged IL cations electrostatically interact with the DNA backbone and grooves.}
\label{fig1}
\end{figure}

Chemical structures of the IL cations and anions, which are used in our study, are presented in Fig.~\ref{fig1}. 

\begin{table*}
\caption{Details of DNA simulated in different IL/water binary mixtures.}
\begin{ruledtabular}
\begin{tabular}{l *{5}{c}} 
 System & Primary solvent component &$[IL]:[water]$ & No. of ILs & No. of water \\
 \hline
$1$ & water & $0:100$ & - & $8000$ \\
$2$ & $[emim][acetate]$ & $80:20$ & $900$ & $2120$ \\
 $3$ & $[emim][acetate]$ & $50:50$ & $500$ & $4719$ \\
 $4$ & $[bmim][acetate]$ & $80:20$ & $900$ & $2475$ \\
 $5$ & $[bmim][acetate]$ & $50:50$ & $500$ & $5497$  \\
 $6$ & $[chol][acetate]$ & $80:20$ & $900$ & $2039$ \\
 $7$ & $[chol][acetate]$ & $50:50$ & $500$ & $4523$\\
 $8$ & $[bmim][lactate]$ & $80:20$ & $600$ & $1900$ \\
 $9$ & $[bmim][lactate]$ & $50:50$ & $500$ & $5499$ \\
 $10$ & $[chol][lactate]$ & $80:20$ & $900$ & $2407$ \\
 $11$ & $[chol][lactate]$ & $50:50$ & $500$ & $5356$ \\
 $12$ & $[bmim][nitrate]$ & $80:20$ & $900$ & $2515$ \\
 $13$  & $[bmim][nitrate]$ & $50:50$ & $500$ & $5578$ \\
 $14$  & $[chol][nitrate]$ & $80:20$ & $900$ & $2072$ \\
$15$& $[chol][nitrate]$ & $50:50$ & $500$ & $4605$ \\
\end{tabular}
\end{ruledtabular}
\label{Table1}
\end{table*}
\section{RMSD of simulated DNA in different IL/water binary mixtures}
\begin{figure}
\centering
\includegraphics{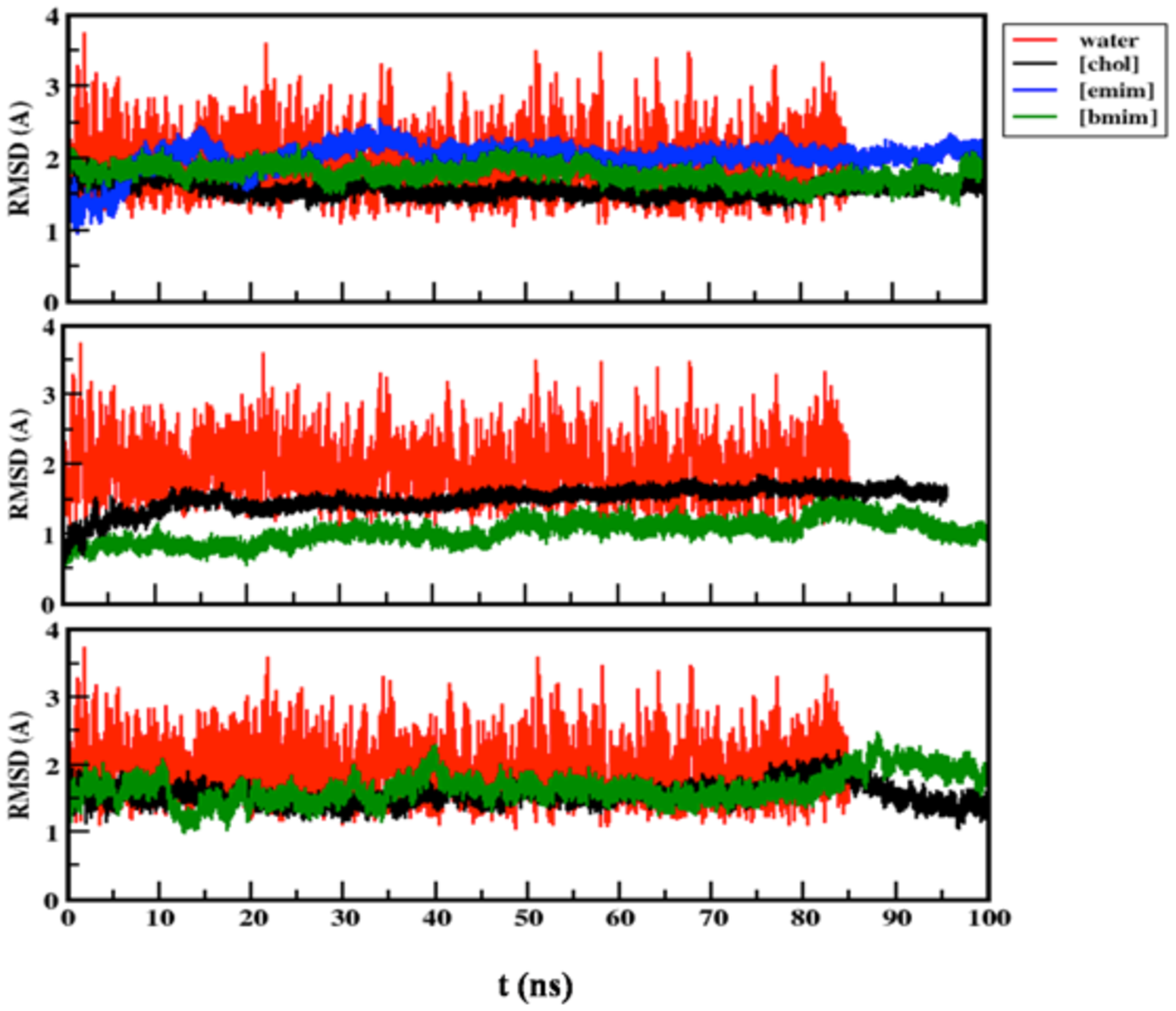}
\caption{Structural characterization of DNA dodecamer. Time evolution of RMSD of simulated DNA in 80:20 IL/water binary mixtures containing the acetate (top panel), lactate (middle panel) and nitrate (bottom panel) anions, relative to the crystal structure.}
\label{fig2}
\end{figure}

\begin{figure}
\centering
\includegraphics{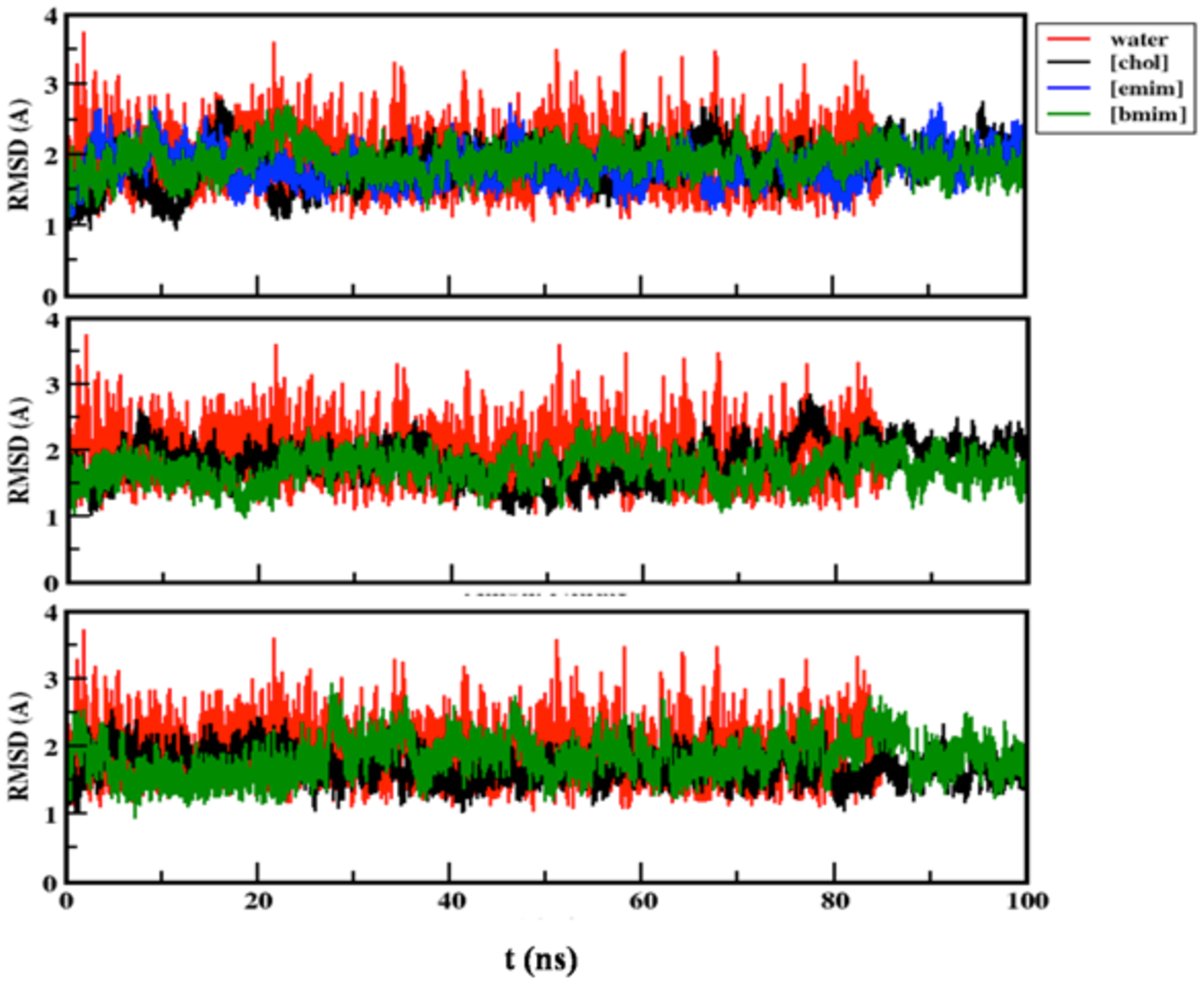}
\caption{Structural characterization of DNA dodecamer. Time evolution of RMSD of simulated DNA in 50:50 IL/water binary mixtures containing the acetate (top panel), lactate (middle panel) and nitrate (bottom panel) anions, relative to the crystal structure.}
\label{fig3}
\end{figure}

Notably structural oscillations of DNA in IL/water systems were dampened with respect to water plausibly due to (i) strong interaction of DNA with IL and (ii) high viscosity of the binary solutions that might restrict DNA dynamics to a certain extent. Another noteworthy feature was that the fluctuations in RMSD (see Figs.~ \ref{fig2}, \ref{fig3}) were relatively lower for DNA solubilized in {\it [imidazolium]}-based ILs, especially in the dilute $50:50$ IL/water solutions. In an earlier study, Senapati et al. \cite{senapati12} reported that the imidazolium cation binds more strongly to DNA than the cholinium cation. The stronger interaction could confer higher rigidity on the imidazolium-bound DNA, leading to the dampened dynamics. Having established the structural stability of DNA in all IL/water binary mixtures, we proceeded to determine the elastic properties of the biomolecule in the different simulated systems.   

\begin{figure}
\centering
\includegraphics[width=0.75\columnwidth]{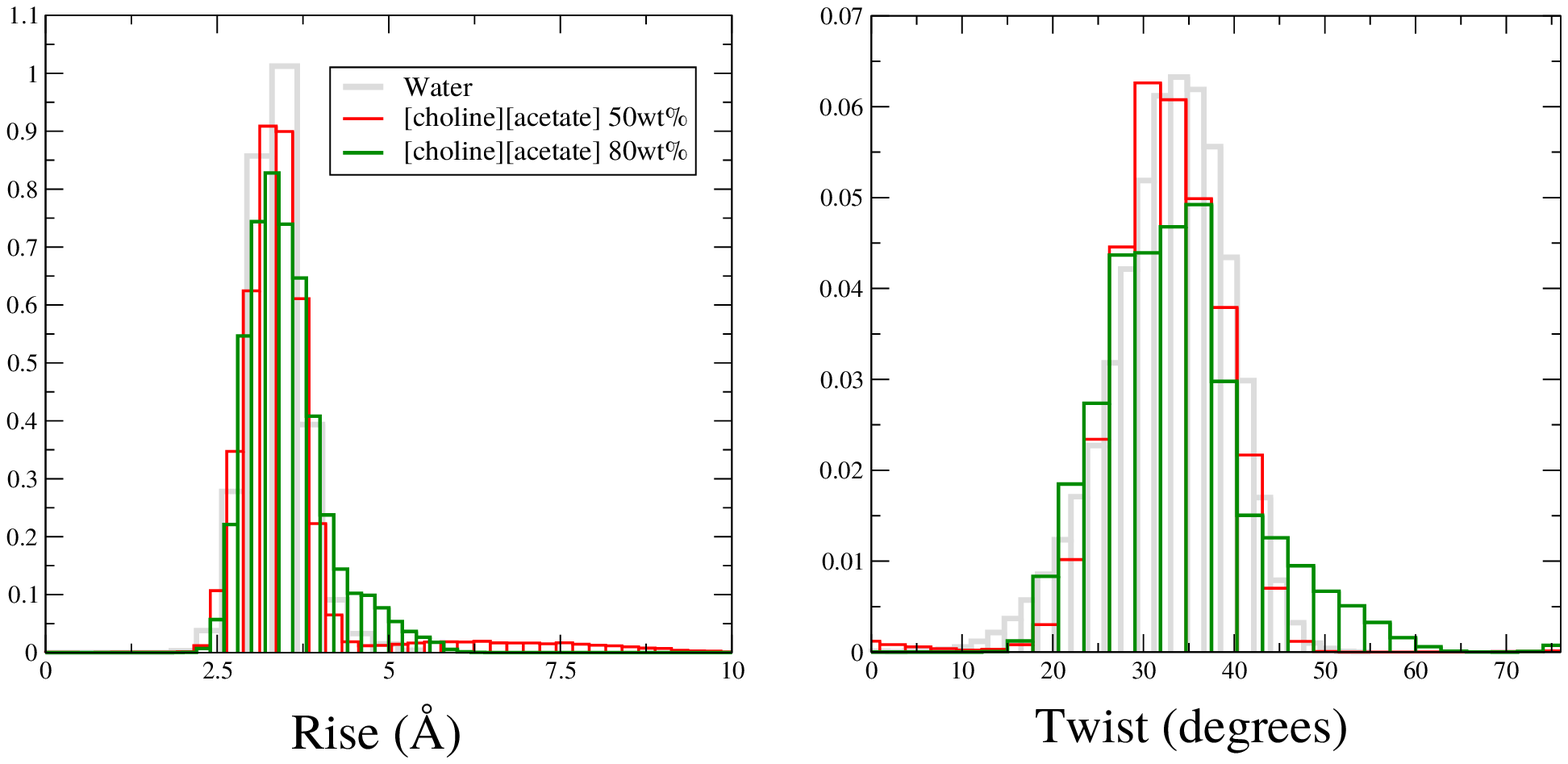}
\caption{Probability distributions of Rise and Twist in \AA{} and degrees, respectively. It depicts that the rise and twist values sampled by the DDd almost equivalent conformational space in presence and absence of [choline][acetate] IL.}
\label{fig4}
\end{figure}

\begin{figure}[tb]
\centering
\includegraphics[width=0.85\columnwidth]{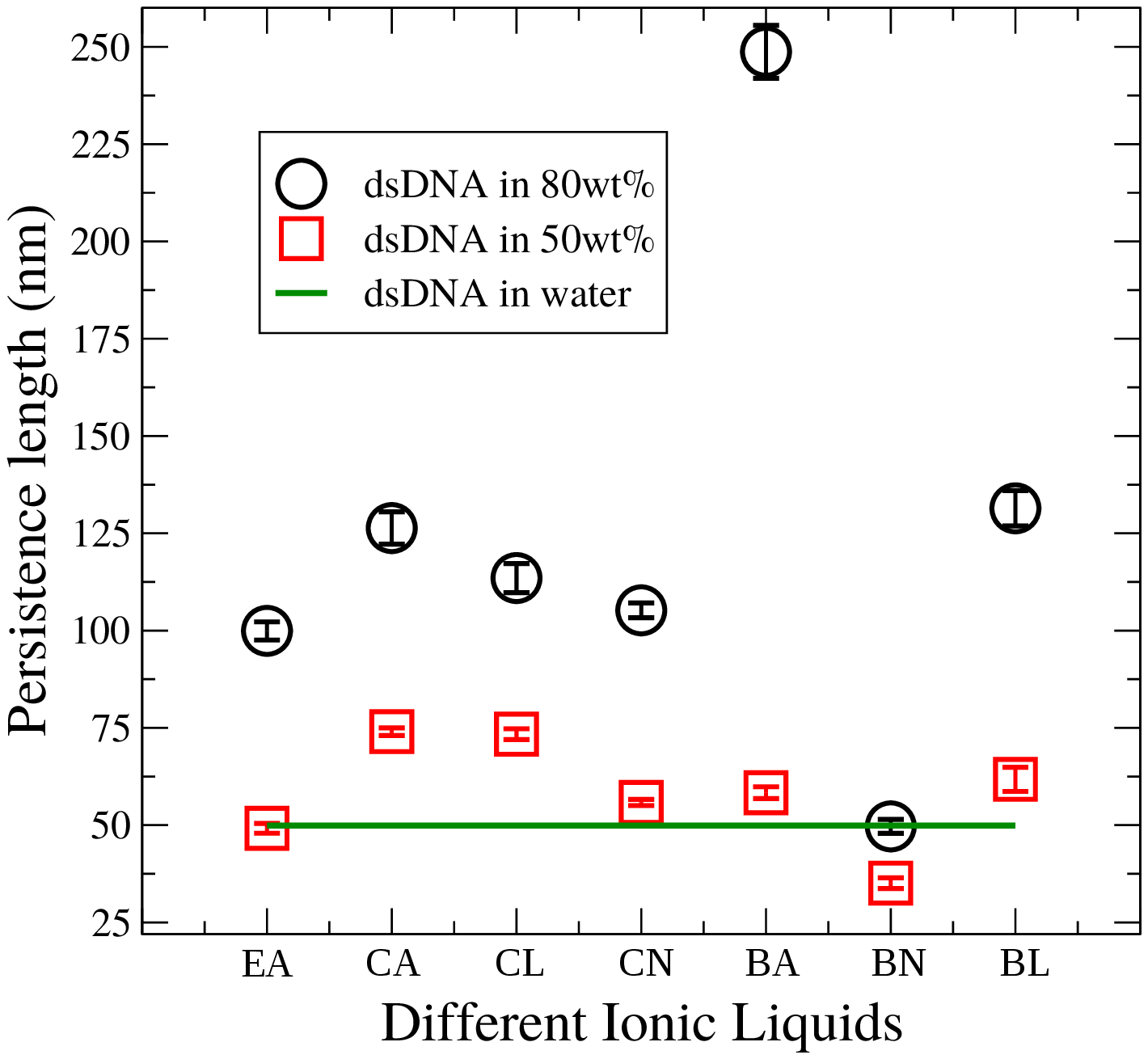}
\caption{Persistence length of DDd calculated for various ILs at different concentrations. EA, CA, CL CN, BA, BN, BL are respectively abbreviated as [emim][acetate], [choline][acetate], [choline][lactate], [choline][nitrate], [bmim][acetate], [bmim][nitrate], and [bmim][lactate].}
\label{fig5}
\end{figure}

\begin{figure}[tb]
\centering
\includegraphics[width=0.85\columnwidth]{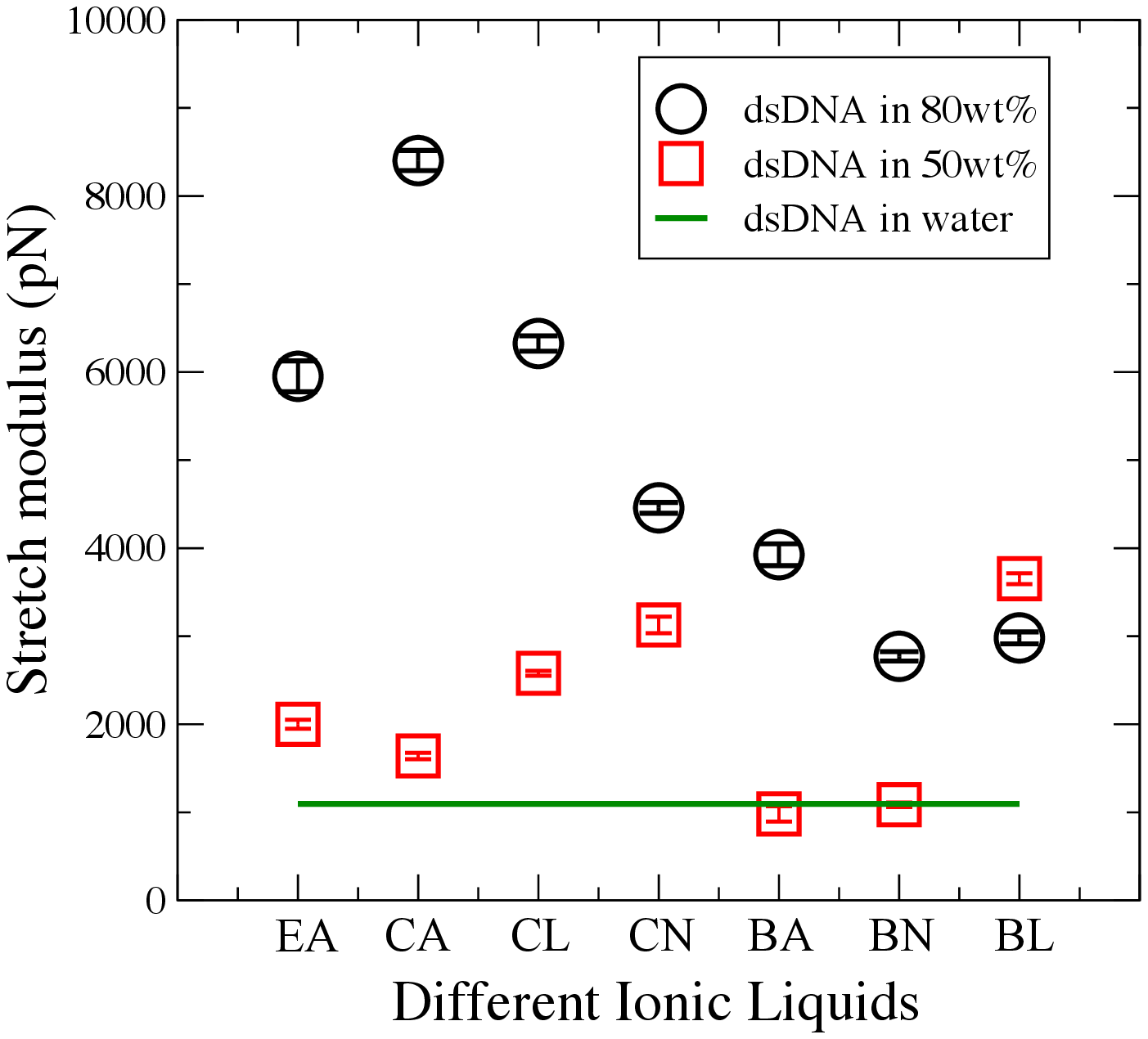}
\caption{Stretch modulus of DDd DNA calculated for various ionic liquids at different concentrations. EA, CA, CL CN, BA, BN, BL are respectively abbreviated as described in Fig. \ref{fig3}.}
\label{fig6}
\end{figure}